# KLOE MEMO 213

# Novel DAQ and Trigger Methods for the KLOE experiment

*The KLOE collaboration*[*].

**Contributed paper n. 224 to XXX ICHEP 2000, Osaka**

## Abstract

KLOE, a new state of the art detector recently commissioned for physics operation at DAΦNE, has many innovative interesting features, especially in the DAQ and Trigger areas. Custom electronics assert a trigger in a 2 microseconds decision time and distributes it to the FEE with a 50 ps time resolution. Data are read out using 10 front-end data acquisition chains and sent to a farm of online servers through a FDDI Gigaswitch. This design of the KLOE DAQ system allows us to manage an input data rate as high as 50 Mbyte/s, and is completely scalable by extending the number of computers connected to the switch.

[*] M. Adinolfi (13), A. Aloisio (7), F. Ambrosino (7), A. Andryakov (6), A. Antonelli (3), M. Antonelli (3), F. Anulli (3), C. Bacci (14), A. Bankamp (4), G. Barbiellini (16), F. Bellini (14), G. Bencivenni (3), S. Bertolucci (3), C. Bini (11), C. Bloise (3), V. Bocci (11), F. Bossi (3), P. Branchini (14), S. A. Bulychjov (6), G. Cabibbo (11), A. Calcaterra (3), R. Caloi (11), P. Campana (3), G. Capon (3), G. Carboni (13), A. Cardini (11), M. Casarsa (16), G. Cataldi (4), F. Ceradini (14), F. Cervelli (10), F. Cevenini (7), G. Chiefari (7), P. Ciambrone (3), S. Conetti (17), E. De Lucia (11), G. De Robertis (1), R. De Sangro (3), P. De Simone (3), G. De Zorzi (11), S. Dell' Agnello (3), A. Denig(4), A. Di Domenico (11), C. Di Donato (7), S. Di Falco (10), A. Doria (7), E. Drago (7), V. Elia (5), O. Erriquez (1), A. Farilla (14), G. Felici, (3), A. Ferrari (14), M. L. Ferrer (3), G. Finocchiaro (3), C. Forti (3), A. Franceschi (3), P. Franzini (9 and 11), M. L. Gao (2), C. Gatti (3), P. Gauzzi (11), S. Giovannella (3), V. Golovatyuk (5), E. Gorini(5), F. Grancagnolo (5), W. Grandegger (3), E. Graziani (14), P. Guarnaccia (1), U. V. Hagel (4), H. G. Han(2), S. W. Han (2), X. Huang (2), M. Incagli (10), L. Ingrosso (3), Y. Y. Jiang (2), W. Kim (15), W. Kluge (4), V. Kulikov (6), F. Lacava (11), G. Lanfranchi (3), J. Lee-Franzini (3 and 15), T. Lomtadze (10), C. Luisi (11), C. S. Mao (2), M. Martemianov (6), A. Martini (3), M. Matsyuk (6), W. Mei (3), L. Merola (7), R. Messi (13), S. Miscetti (3), A. Moalem (8), S. Moccia, (3), M. Moulson (3), S. Mueller (4), F. Murtas (3), M. Napolitano (7), A. Nedosekin (3 and 6), M. Panareo (5), L. Pacciani (13), P. Pages (3), M. Palutan (13), L. Paoluzi (13), E. Pasqualucci (11), L. Passalacqua (3), M. Passaseo (11), A. Passeri (14), V. Patera (3 and 12), E. Petrolo (11), G. Petrucci (3), D. Picca (11), G. Pirozzi (7), C. Pistillo (7), M. Pollack, (15), L. Pontecorvo (11), M. Primavera (5), F. Ruggieri (1), P. Santangelo, (3), E. Santovetti (13), G. Saracino (7), R. D. Schamberger (15), C. Schwick (10), B. Sciascia (11), A. Sciubba (3 and 12), F. Scuri (16), I. Sfiligoi (3), J. Shan (3), P. Silano (11), T. Spadaro (11), S. Spagnolo (5), E. Spiriti (14), C. Stanescu (14), G. L. Tong (2), L. Tortora (14), E. Valente (11), P. Valente (3), B. Valeriani (10), G. Venanzoni (4), S. Veneziano (11), Y. Wu (2), Y. G. Xie (2), P. P. Zhao (2), Y.Zhou (3)

(1) Dipartimento di Fisica Universita e Sezione INFN, Bari, Italy.
(2) Institute of High Energy Physics of Academica Sinica, Beijing, China.
(3) Laboratori Nazionali di Frascati INFN, Frascati, Italy.
(4) Institut für Experimentelle Kernphysik, Universität Karlsruhe, Germany.
(5) Dipartimento di Fisica Universita e Sezione INFN, Lecce, Italy.
(6) Institute for Theoretical and Experimental Physics, Moscow, Russia.
(7) Dipartimento di Scienze Fisiche Universita e Sezione INFN, Napoli, Italy.
(8) Physics Department, Ben-Gurion University of the Negev, Israel.
(9) Physics Department, Columbia University, New York, USA.
(10) Dipartimento di Fisica Universita e Sezione INFN, Pisa, Italy.
(11) Dipartimento di Fisica Universita e Sezione INFN, Roma I, Italy.
(12) Dipartimento di Energetica dell'Universita, Roma I, Italy.
(13) Dipartimento di Fisica Universita e Sezione INFN, Roma II, Italy.
(14) Dipartimento di Fisica Universita e Sezione INFN, Roma III, Italy.
(15) Physics Department, State University of New York at Stony Brook, USA.
(16) Dipartimento di Fisica Universita e Sezione INFN, Trieste/Udine, Italy.
(17) Physics Department, University of Virginia, USA.)

1 Introduction.

The main goal of the KLOE experiment [1] at the DA$\Phi$NE $e^+e^-$ collider in Frascati is to measure CP violation parameters in the Kaon system with a sensitivity of one part in ten thousand.

The KLOE detector consists of three main parts; a large cylindrical tracking chamber DC [2]; a hermetic lead-scintillator fibers electromagnetic calorimeter EmC [3]; a large magnet surrounding the whole detector, consisting of a superconducting coil and a iron yoke. DA$\Phi$NE peak luminosity is expected to be $10^{33} \text{cm}^{-2}\text{s}^{-1}$, which corresponds to 10 kHz of estimated data rate with an average event size of 5 kbytes. About half of it is due to the $\Phi$ decays while 5 kHz are expected to be due to prescaled Bhabha and cosmic events. Given the very high event rates, and the extreme target accuracy in the measurement of $\varepsilon^|/\varepsilon$, the requirements for the trigger and DAQ system are very stringent and somewhat unprecedented in $e^+e^-$ collider experiments. The trigger system in fact has, first of all, to be as close as possible to 100% efficiency on both the charged and the neutral decay of $K_l K_s$, but with a good background rejection. Secondly it has to produce a valid signal within 350 ns from the bunch crossing, in order to start the Electro magnetic Calorimeter ADCs [4] and TDCs [5] conversion, Moreover a fixed dead time of about 2 $\mu$s must follow each trigger to allow the drift mechanism in the chamber cells to take place. Last but not least all the relevant information provided by the trigger system must be acquired in order to monitor the stability and the behaviour of the trigger itself.

DAQ requirements are as challenging as the trigger ones. In fact the DAQ system must be able to sustain 50 Mbytes/s, be scalable and flexible, asynchronous with respect to the trigger signal, have a fixed dead time, and be optimised for low channel occupancies. To keep up with these demands, the trigger and DAQ systems have been designed as a mixture of custom specialised and high speed commercial electronics. Moreover an efficient software architecture has been designed and developed to deal with this complex environment.

2 Overview of the trigger and DAQ systems.

The KLOE trigger system [6], [7] is based on local energy deposits in the EmC and hit multiplicity information from the drift chamber.

A two level scheme has been adopted in order to produce a first early trigger signal (T1) to start EmC FEE digitisation while the second level trigger signal (T2), delayed by 2 μs with respect to T1 is designed to fully exploit the information coming from the drift chamber. Figure 1 shows the trigger logical layout.

Specifically, after the arrival of a first level trigger, additional information is collected from the drift chamber, which is used, together with the calorimetric information, to confirm the first level, to start digitisation of the drift chamber FEE [8] and to start the DAQ read out.

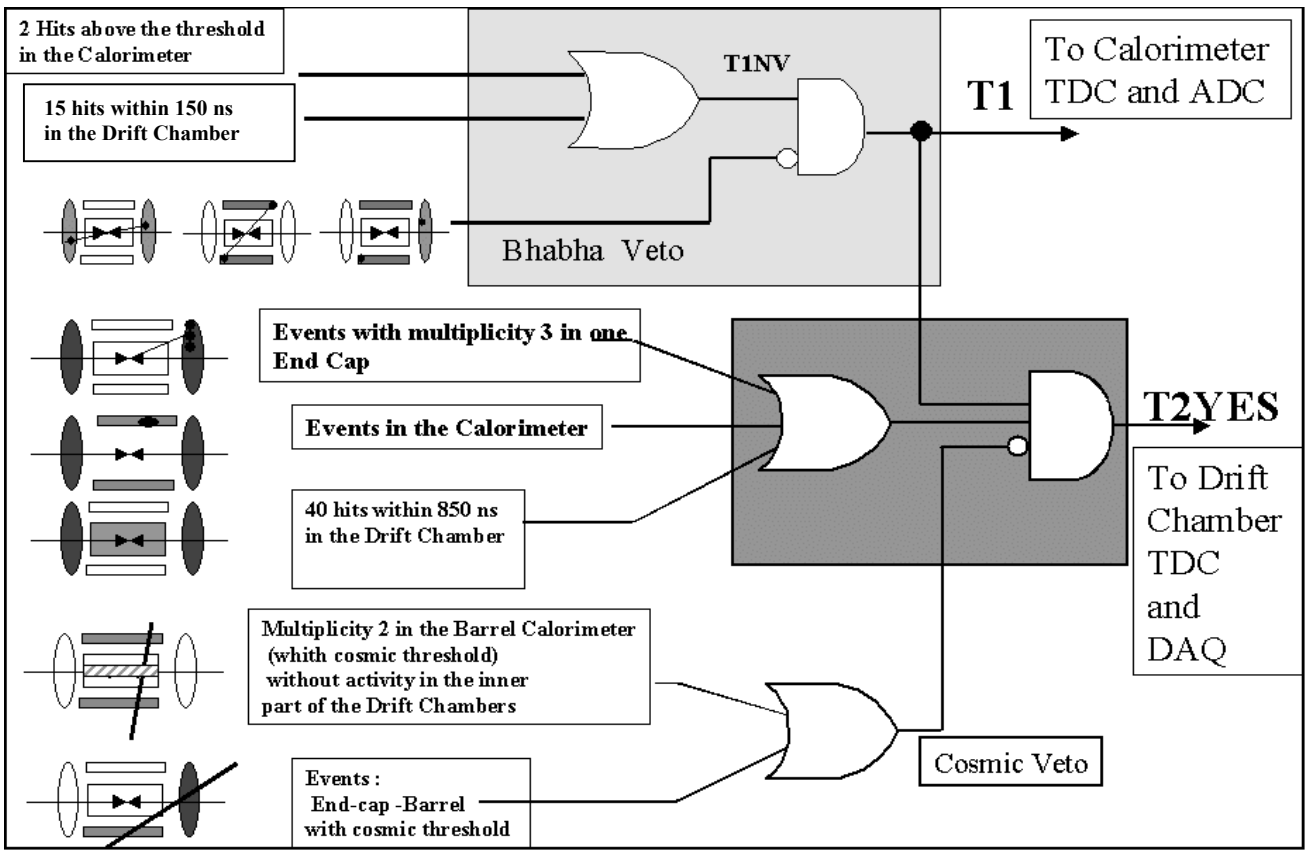

Fig. 1: Trigger logic.

The time between bunch crossings in DAΦNE is 2.7 ns, therefore the trigger must operate in continuous mode. Signals from EmC and the DC are properly treated to generate two separate trigger decisions, that are OR-ed to form the final T1 and T2 signals. To avoid the intrinsic jiitter of the trigger signal formation the first level trigger is synchronised within 50 ps with the machine radiofrequency before being delivered to the FEE. Nevertheless the bunch crossing originating the event has to be determined by the off-line analysis.

The trigger hardware consists of 13 different types of custom-designed electronics modules, needed either to generate the elementary trigger signals from the various subdetectors or to produce the final trigger signals to be delivered to the FEE modules.

The data taking process is event-driven [9] and built around custom protocols and read-out controllers. Each controller hosts a trigger FIFO to store the information of the incoming trigger and a memory to store the event fragment coming from FEE. Therefore the data taking process is asynchronous with respect to the second level trigger signal.

3 Trigger hardware architecture.

As previously underlined the trigger hardware architecture exploits both the signals coming from EmC and the ones coming from DC. The trigger architecture is therefore divided in two chains one for the EmC, one for the DC. Their signals are OR-ed to produce the trigger signal.

3.1 Calorimeter trigger chain.

The signals from the calorimeter's PM's are collected by 164 boards (SDS/Splitter boards) and are split in three different paths, through which the signals are delivered to the ADCs, TDCs and trigger boards respectively. In the trigger path five adjacent PM signals corresponding to one calorimeter column are summed, for each side of the calorimeter separately.

In addition, groups of 6 adjacent PM signals in the outmost layer of the calorimeter are summed in order to form the signals for the cosmic veto.

The trigger sector signals are formed in the "PIZZA" by grouping together the output signals of the Splitter. Each "PIZZA" output defines a topologically connected region called sector. The analog signals corresponding to each trigger sector are compared to a set of different thresholds and processed in a logic state machine.

This task is performed by the "Digitiser-Shaper (DISH)" module. DISHes provide output signals indicating local energy deposits above a certain threshold in the calorimeter. The number of hits in the calorimeter is determined in the "Precise Analogue STAge (PASTA)" which delivers the information to the "Trigger Organiser and Timing Analyser (TORTA)" for the final decision. Furthermore all the information provided by DISHes and the ones provided by the TORTA are acquired and allow us for continuous monitoring of the detector and the trigger system.

3.2 Drift Chamber trigger chain

The FEE stage of the drift chamber is made of 280 "Amplifier Discriminator Shaper (ADS)" boards, which provide digital signals for the TDCs and the signals for the trigger. To achieve this purpose, the DC signals are first shaped to a width of 250 ns, then are summed by the ADS boards in groups of 12 contiguous wires, and finally sent to 20 "Sum Unit Providing Plane Information (SUPPLI)" boards, housed in the same FEE crates, where the counting of the fired wires on half drift chamber plane is performed. The output signals from the SUPPLI boards are sent to three PreCaffe boards, which join the half-plane multiplicity information and pack the drift chamber layers in groups of 4-6. These modules produce as an output signal the multiplicity from the 10 DC superlayers. At the end of the chain the "Chamber Activity Fast Fetch (CAFFE)" board sums the signals of the superlayers to produce an output in current within 250 ns which is proportional to the number of fired wires in the chamber.

A first level drift chamber trigger (T1D) is issued whenever this current exceeds a given programmable threshold. This signal is also integrated during the following 1 µs and a second level signal (T2D) is issued if another programmable threshold is exceeded. Again all the information coming from the CAFFE is acquired.

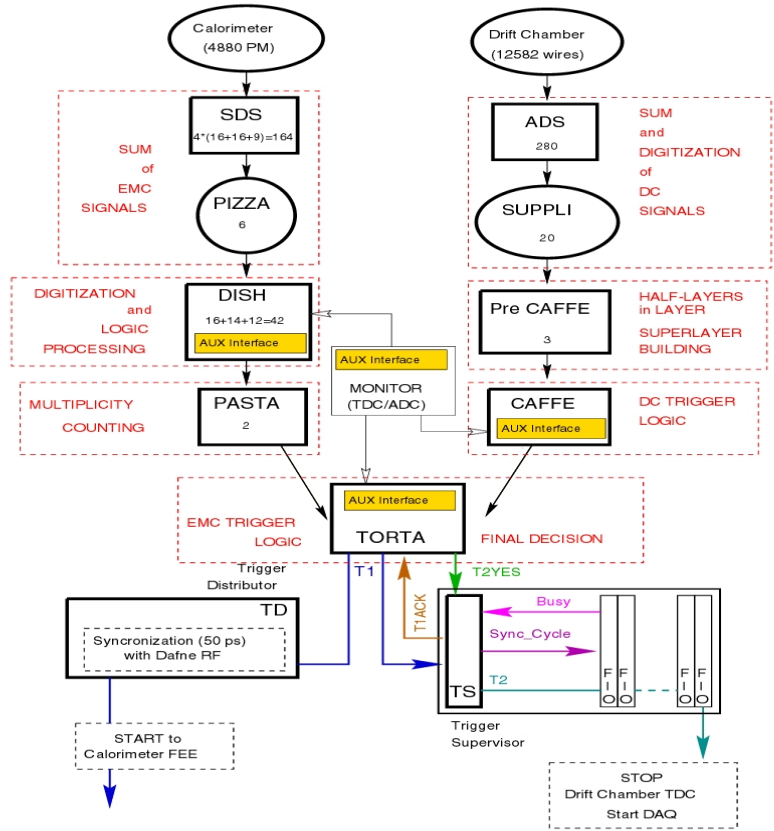

Fig. 2: Trigger hardware layout.

## 3.3 The trigger logic box and the trigger DAQ interface

The multiplicity signals from the calorimeter and the first and second level DC triggers, T1D and T2D, are sent to the TORTA, that merges all the information and on the basis of programmable logic tables takes the final decision and delivers the trigger signal. The TORTA distributes T1 to the Trigger Supervisor (TS) and receives back the T1ACK signal which disables further T1 generation for a fixed time (2.6 µs).

The T1 is also delivered to the Distributor module (TD), which performs the synchronisation with the machine RF and distributes the T1 to the EmC FEE. Whenever the TORTA generates a T2yes (within 1.5 µs from the T1) the TS asserts a T2 1.8 µs after T1. The signal is sent to the "Fan in-out (FIO)" modules, which distribute it to the corresponding readout chain with a precision of 1 ns. Special trigger cycles have been implemented in the TS in order to check that all FEE and read-out controllers received the same number of triggers. Whenever this cycle is issued by the TS a Sync signal is asserted and all the FEE deliver to the read-out processors the last trigger number received.

The read-out processors then check the alignment between their own trigger number and the trigger number received by the FEE. Should a misalignment occur an halt signal in asserted and the DAQ process stops.

## 4 DAQ hardware architecture.

One important requirement to the DAQ system is scalability.
To reach this purpose the system is divided in ten modular structure of interconnected VME crates. Four VME chains are devoted to the acquisition of the EmC, four to the DC and two to the acquisition of the trigger. These chains are made of up to six crates, each with sixteen slave boards and a local read-out controller –the ROCK [10]. All Rocks in the chain are interconnected to a controller manager, ROCKM [11] residing in a VME crate together with a VME processor board (DEC EBV14).

A VIC bus interconnection scheme allows the VME CPU to address all the VME boards installed in the chain. The ROCK performs crate level read-out and gathers data from the slaves using the AUX-bus, a custom protocol developed to enhance the event-driven behaviour of the KLOE DAQ. With a single broadcast transaction (trigger cycle), the AUXbus allows the crate controller to find the DAQ boards with valid data for a specific event number. Data transfer is then carried out using high-speed random length block transfers, with an asynchronous VME-like handshake (data cycle). The ROCK then builds data frames consisting of an event number, slaves data and a parity word. In the same fashion, the ROCK Manager performs chain level read-out, collecting the data frames belonging to a given event number from all the ROCKs. In order to extend the event-driven behaviour at the chain read-out, the Manager implements an AUXbus companion protocol, the Cbus.

Similarly to the AUXbus, the Cbus tags data transactions with event numbers, making the entire chain an event-number-driven machine. After a system initialisation phase the data taking process is fully handled by the ROCKs and the ROCK Manager and does not require any CPU activity on the FE boards. For each event number, the ROCK Manager assembles a common data frame containing the respective data from all the ROCKs in its chain. In this way, each chain shares also the functionality of a sub-event builder.

The VME processor is then in charge of moving the frames on the VMEbus to an online farm of 4-ways H50 IBM servers. Sub-streams of data containing the same trigger number coming from the detector must be delivered to the same farm component to perform final event-building and checks.

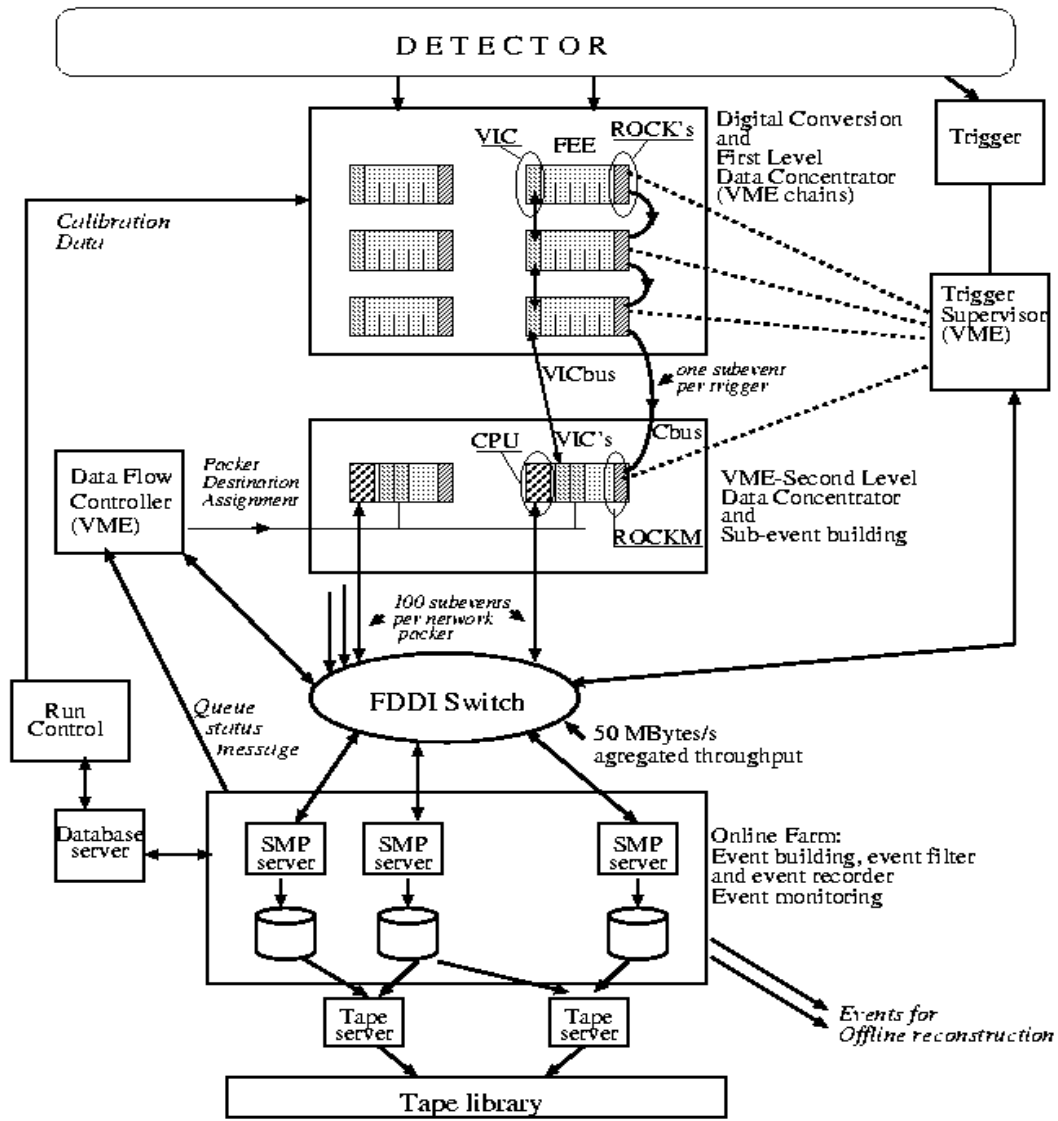

Fig. 3: DAQ stream

A Data Flow Controller (DFC) process has been developed to perform this synchronisation task. All the VME processors share the DFC tables of trigger-farm assignment via an horizontal VIC-bus connection which interconnects the ten chains. Each chain is connected to the farm via an FDDI link.

Data are finally archived in an IBM3494 tape library.

5 DAQ software architecture.

Many specialised tasks running on a multi-platform computing environment are the core of the KLOE software architecture. These tasks are responsible of taking care and checking the full data flow, from VME front-end electronics down to tape recording devices. As described before, the first levels of data concentration are handled by VME read-out controllers and do not require any CPU activity but for the initialisation procedure. In this phase, thresholds, trigger settings and run conditions, are downloaded into the VME CPUs. Initialisation data are memorised in a DB2 database [12] where other KLOE and DAΦNE conditions are also registered during the run. Once framed in the ROCK Manager memory, streams of sub-events corresponding to a set of triggers are read-out by the VME processor using block transfers (BLT). The BLT cycle length is dynamically optimised according to the effective throughput of the chain. Two software processes – the Collector and the Sender [13]– which run asynchronously with respect to each other on the VME processor manage the read-out activity.

The Collector reads ROCK Manager memory and pushes the data frames in a shared memory with a FIFO structure.

The Sender retrieves from the queue a predetermined amount of sub-events and dispatches them to the on-line computer farm via an FDDI optical link using the TCP/IP protocol. The information allowing stream synchronisation is taken from Data Flow Control (DFC)[14]. This task, which runs on a dedicated HP VME processor, implements a modified round-robin protocol to assign the right farm address to packets of events.

Trigger Supervisor and FIOs are controlled by a dedicated task (L2trigger) running in their VME processor. This task is in particular responsible for managing error conditions.

Any time an halt signal is received in ths TS, originated by the FEE, an error diagnosis logic flow is triggered and, by interrogating the FIOs, chain and crate are identified. An interrupt is delivered to the Collector process which manages that chain and this process starts a deep diagnostic procedure which leads us to identify the source of the problem.

The data sub-streams are delivered to the farm processors through FDDI links. Since the data acquisition system has been partitioned in ten different chains all the sub-streams of data must be rebuilt.

This happens soon after receiving the ten different data streams by means of a process the Builder, which is decoupled from the one which receives the data, – the Receiver, by a shared memory with ten FIFO structures. Once data frames are properly rebuilt on each farm processor they are written on disk using Ybos format and later archived.

The use of shared memories to decouple all tasks cooperating in the event data flow, allowed us to implement on line event-spy processes in order to monitor detector, machine conditions and physics, without disturbing the main data flow. Specialised software libraries, "the KLOE integrated Dataflow KID" [12] allow us for distributing spy processes into different servers, according to the needed computing resources. All DAQ processes, that are co-ordinated by a central run-control, can communicate between them using the KLOE message passing system [12].

6 Slow-control

The status and all the relevant parameters of the KLOE detector are monitored by the Slow-control system, that is fully integrated in the framework of the DAQ; this allows to keep track of all the information making them available at the level of the event building (as shown in Fig. 4) and it is the first time it is done in a large experiment.

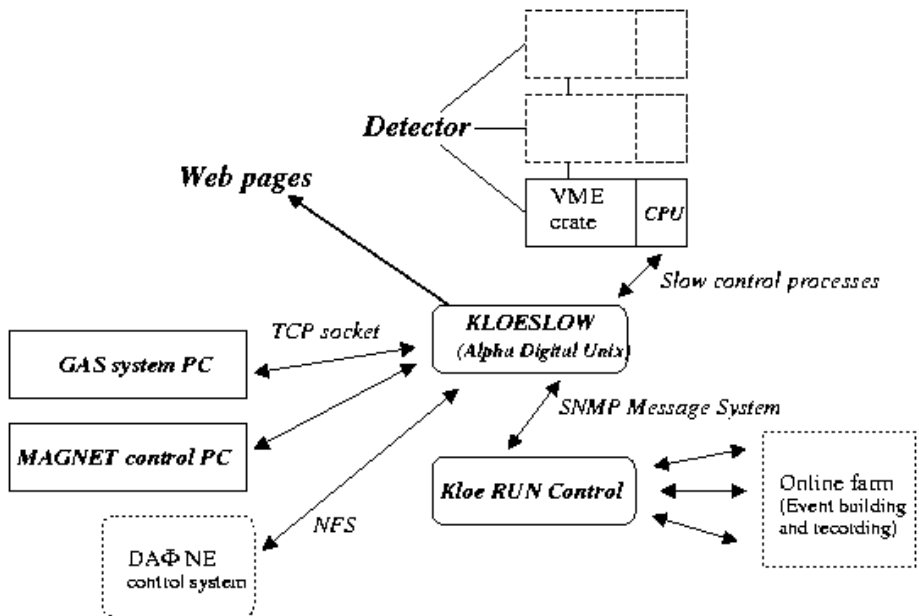

Fig. 4: Slow-control schematic layout and data flow.

The sharing of data between the KLOE and the DAΦNE control systems completes such a strong integration.
A Digital Unix Alpha machine and on one VME crate of the DAQ system with a standard CPU and six CAENet serial interface modules build up the system that controls all the parameters of the FEE crates, as well as the high and low voltages of the detector: the ten DAQ chains, the ADS crates for the low voltage of the drift chamber, the SDS crates for the low voltage of the calorimeter, the high voltage both for the drift chamber and the calorimeter. Different processes running on the VME processor board take care of the monitoring and setting of all parameters for crate control and read-out and the inter-process communication takes advantage of the KLOE message passing system. The user interface is entirely made in HTML technology: a Web server running on the Alpha machine can display different pages; the monitoring and setting of the parameters is done using CGI programs.
The Slow-control also monitors parts of the detector not strictly embedded in the online structure such as the control systems of the drift chamber gas and of the superconducting coil of the KLOE magnet, handled by remote PCs. Both communicate with Slow-control processes through the TCP/IP protocol.

The accelerator conditions are also available thanks to the data sharing between the KLOE and DAΦNE control systems.

The Web interface is also used to display other relevant information such as the measurements coming from the event filtering and monitoring tasks (see below): luminosity, beam position and momenta, etc.

The Slow-control is strictly integrated with the Run-control through the shared memory and the message system: on one end the Slow-control data are shared to the event-building and recording processes of the online farm, on the other the Run-control can trigger all the initialisation procedures of the Slow-control (such as threshold settings, channel masking or veto) at the start of the acquisition. In this way is possible to store the data about the status of the detector and of the accelerator event-by-event or in the run header (the slow varying ones) in the Ybos format, making them easily accessible in the analysis phase. Those data are also logged to the general DB2 database of the experiment.

7 Event Filtering and Monitoring tasks

Using the shared memory filled by the farm Builder, the YBOS raw event structure is available to many other processes. The Trgmon process, for example, uses the pattern of reconstructed trigger sectors to provide a fast monitor of relevant quantities such as instant luminosity, background levels, data rates. The L3 process is instead based on a full reconstruction of calorimeter cells, timing position and energy. The L3 creates shared memories of three categories: Bhabha and $\gamma\gamma$ events, Minimum ionisation particles and cosmic rays. These events are used by the online calibration and also permit a fast powerful monitor of the detector and the event reconstruction behaviour. Both processes relies on a specialised set of "C" routines that perform partial "event" reconstruction at an element level, both for the trigger and the calorimeter.

An Analysis Control (A_C) [15] based set of programs, grouped under the "Physmon" task can get connected to the three shared memories and perform special operations. Just as an example, the A_C module "Trkmon" allows us to calculate for each run the average Bhabha momentum, beam position and beam size, together with a reliable control of hardware and software drift chamber wires efficiencies. Similarly, "Calmon" and "Survey" allow us a tight control of calorimeter and trigger reconstruction.

Histogram-server and Event-display [16] fetch event from one of the above described shared memories. Dedicated histogram browsers check for "dead" detector channels and monitor the general behaviour comparing with reference histograms.

8 On beam operation and conclusions

The KLOE Trigger and DAQ system underwent a period of data taking from June to December 1999 and 2.4 pb$^{-1}$ were collected.

The DAQ system worked efficiently all over the data taking period.

The typical sustained data acquisition rate was 900 Hz, among which 700 Hz are unvetoed cosmic about 200 Hz machine background and unvetoed Bhabha and a few Hz of ϕ. Peak rate as high as 10 kHz were tested just lowering thresholds in the FEE boards. The synergy between the Trigger and the DAQ within the above described architecture, allowed us to monitor the behaviour of the detector and at the same time to measure luminosity and ϕ rates on line.

Figure 5 presents an example of the use of CPU power in one server of the online farm for different requirements of data throughput to be sustained by the single server. The process involved in the data acquisition are well optimised. Peaks in data throughput corresponding to the maximum FDDI capability of 12.5 MB/s have been supported, requiring about 50 % of the server resources. By dispatching data to different servers throughput in excess of 50 MB/s have been observed; this is the highest throughput obtained by a collider experiment so far. Monitor's processes, like Trgmon and L3filter are limited to use only one processor, therefore no busy conditions in the trigger system are originated because of lack in server resources.

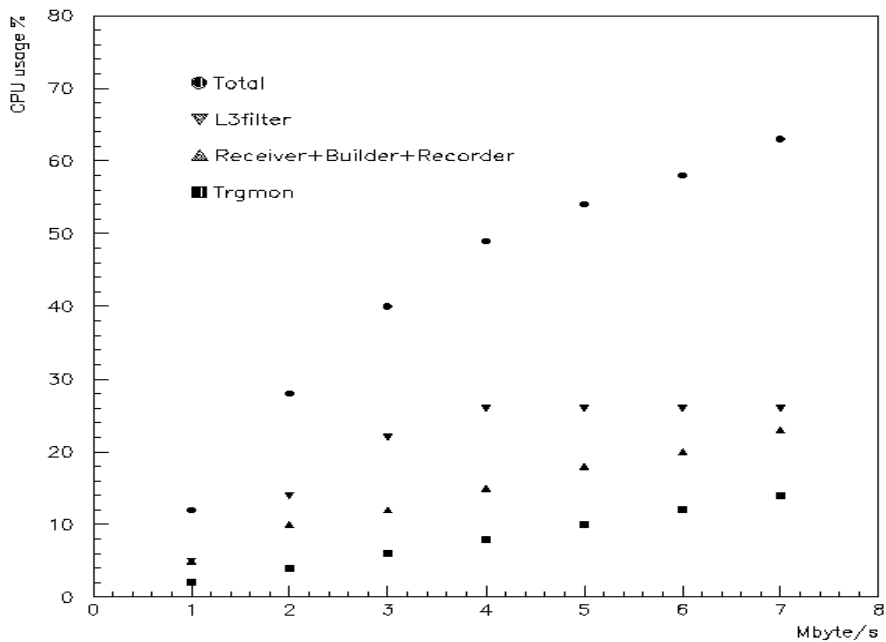

Fig. 5: CPU usage.

Trigger system redundancy allowed us to measure trigger efficiency on different physics channel from the data themselves and on relevant ϕ decays were of the order of 99 %.

For the future we foresee to introduce on line DC and EmC calibration by using cosmic tagged and downscaled by the Trigger system.